# Preparation and characterization of magnesium coating deposited by cold spraying

Xinkun Suo [a, *], Xueping Guo [a], Wenya Li [b], Marie-Pierre Planche [a], Rodolphe Bolot [a], Hanlin Liao [a], Christian Coddet [a]

[a] *LERMPS, Université de Technologie de Belfort-Montbéliard, Site de Sévenans, 90010 Belfort Cedex, France*

[b] *Shaanxi Key Laboratory of Friction Welding Technologies, Northwestern Polytechnical University, Xi' an 710072, Shaanxi, China*

\* Corresponding author: Xinkun Suo

Address: LERMPS, Université de Technologie de Belfort-Montbéliard, Site de Sévenans, 90010 Belfort Cedex, France

Tel.: +33 03 84 58 37 36

Fax: +33 03 84 58 32 86

E-mail address: suoxinkun@gmail.com

**Abstract**

Magnesium (Mg) and its alloys have a great potential as structural materials due to their beneficial combination of high strength to weight ratio, high thermal conductivity and good machinability. However, few literatures about Mg coatings fabricated by cold spraying can be found. In this study, Mg coatings were fabricated by cold spraying, and the microstructure, phase structure, oxygen content and microhardness of the coating prepared under different main gas temperatures were investigated. The critical velocity of the particle was evaluated through numerical simulations. The particle deformation behavior and bonding mechanism were discussed. The result of the oxygen content test shows that the oxygen contents of the coatings did not increase comparing with that of the feedstock powder. The simulation results show that the critical velocity of Mg particles was in a range of 653 m/s to 677 m/s. The observation of the coating fracture morphology shows that the formation of the coating was due to the intensive plastic deformation and mechanical interlocking. The microhardness of the coating increased with the increase of the main gas temperature from 350$^\circ$C to 450$^\circ$C due to the decrease of the coating porosity.

# 1. Introduction

Mg and its alloys possess a great potential as structural materials due to their highest strength to weight ratio among all metallic structural materials which could thereby contribute to a wide range of applications, such as the automotive industry, computer parts, aerospace components, mobile telephones, sport equipments, handheld and household tools (Cao et al., 2006; Yang et al., 2008). Meanwhile Mg is a very active electrochemical metal and can generate an anode reaction with other metallic materials (Carboneras et al., 2010; Gonzalez-Nunez et al., 1995), which makes Mg and its alloys have a special potential for the application as anode protective coatings in many corrosive media. In addition, Mg and its alloys have a potential as biomedical materials due to their good biocompatibility and good machinability (Zhang et al., 2009; Zhang et al., 2010). Casting is a principal process to produce Mg and its alloys, and this process acquires limited success (Francis and Cantin, 2005; Xiao et al., 2009). Due to the long dwelling time at a high operating temperature, the oxidation of Mg is inevitable. In order to make use of the advantage of Mg and its alloys, it is necessary to develop another effective process capable to prevent Mg and its alloys from oxidization.

Conventional thermal spraying technologies are not suitable for fabricating the coatings of Mg and its alloys for the raison of its strong chemical activity at high temperature and in fusion state. As an advanced coating process, cold spray technology, also called cold gas dynamic spray (CGDS) or kinetic spray, employs normally a De-Laval type nozzle and high pressure thermal gas to accelerate powder particles to a high velocity (300–1200 m/s), and then deposits them onto a substrate to form a coating as a result of the intensive plastic deformation of particles. Depending on the velocity, the particles will either bond to the substrate or erode the substrate. The velocity at which bonding is achieved is referred as the critical velocity, which is directly dependent of the natures of the particle and substrate. The bonding mechanism is assumed to be a result of the adiabatic shear instability at the interface during impact which occurs as a result of high strain rate deformation processes (Gilmore et al., 1999; Hussain et al., 2009; Van Steenkiste et al., 2002; Van Steenkiste et al., 1999).

Most metals and alloys with good plastic deformation ability, such as Fe, Al, Ni, Cu and some of their alloys, have been widely investigated. Typically critical velocities of 620–640 m/s, 680–700 m/s, 620–640 m/s and 560–580 m/s for Fe, Al, Ni and Cu powders have been reported respectively (Moy et al., 2010). Moreover metallic matrix composites have been fabricated using cold spraying due to the plastic deformation of metallic matrix, such as WC-Co (Ang et al., 2011), SiC-metals (Sansoucy et al., 2008), $B_4C$-metals (Yandouzi et al., 2010). Some metals with poor plastic deformation ability due to their close-packed hexagonal structure can also be prepared by cold spraying, such as Ti and its alloys (Moy et al., 2010). Furthermore, owning to the low process temperature (lower than the melting point of the sprayed material) and very short heating dwelling time (less than 1 ms), cold spraying can be applied to deposit heat-sensitive and oxidation-susceptible materials, for instance of Zn, Al and their alloys (Lee et al., 2008; Morimoto et al., 2004).

Mg belongs to the hexagonal system and has few slip systems, and the plastic deformation of Mg particle is quite difficult. To date, coatings of Mg and its alloys prepared by cold spraying have been little reported. In this study, Mg coatings were fabricated by cold spraying in atmospheric environment. The critical velocity of Mg particle was evaluated by numerical simulations. The mechanisms of deformation and bonding of Mg particle were discussed.

## 2. Experimental

### 2.1. Materials

A commercial pure Mg powder used in this study was produced by Fusen Magnesium Powders Co. Ltd, Nanyang, China. Figure 1 shows its morphology and etched cross-sectional microstructure. It can be found that the powder particles present an irregular shape (as shown in Fig. 1(a)) and equiaxed grain structure (as shown in Fig. 1(b)). The particle size distribution was characterized by a commercial system MASTERSIZER 2000 (Malvern Instruments Ltd., UK). Figure 2 shows that the particle size distribution is in a range of 30 µm to 120 µm, and the average particle size is 63.2 µm.

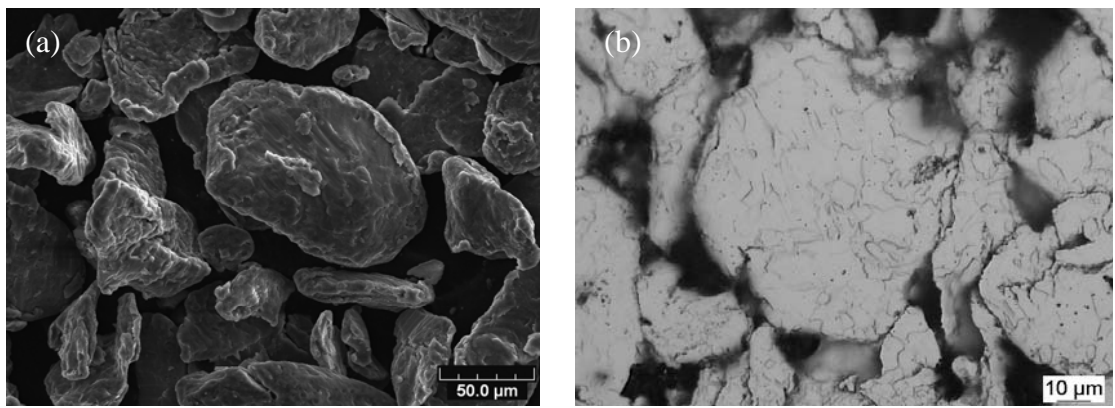

Fig. 1. Morphology of the used powder particles: (a) surface, (b) etched cross-section.

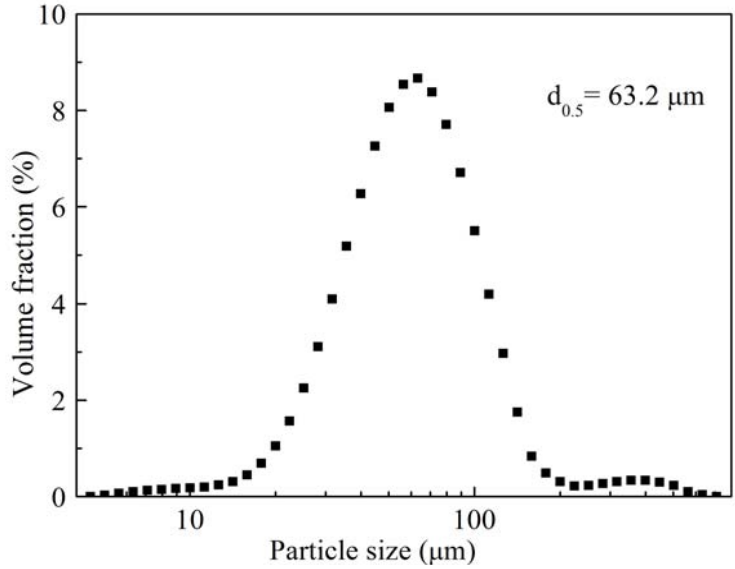

Fig. 2. Size distribution of the used powder particles.

**2.2. Numerical simulation modeling**

Numerical simulations were performed using commercial software FLUENT (Ver. 12). The details of the dimension and scheme related to the spray nozzle and boundary conditions can be found in many literatures (Grujicic et al., 2004; Li et al., 2007; Takana et al., 2008). The particle diameters were set as 30, 40, 60, 80, 100 and 120 μm, and the shape factor of the particle was 0.9 due to its irregular shape.

**2.3. Coating deposition**

A cold spray system with a commercial spray gun (CGT GmbH, Germany) was employed to deposit coatings. Compressed air and argon were used as the main gas and powder carrier gas, respectively. The pressure of the main gas at the pre-chamber was 2.5 MPa, and the temperatures of the main gas were 300, 350, 400, 450, 500 and 630$^\circ$C for different trials. The powder feed rate was 70 g/min. The standoff distance from the nozzle exit to the sample surface was 30 mm. The traverse speed of the gun was 100 mm/s. Low carbon steel plates were used as substrates.

**2.4. Characterization of the powder and cold-sprayed coatings**

The surface morphology of the starting powder was observed using scanning electron microscope (SEM, JEOL, JSM-5800LV, Japan) in the mode of the secondary electron image. The cross-section microstructures of the coatings were examined using SEM in the mode of the scattered electron image. The grain structures of the starting powder and as-sprayed coatings were inspected by optical microscope (OM, Nikon, Japan). The polished cross-sections of the coatings were etched by a solution composed of 4 ml $HNO_3$ + 100 ml alcohol. The porosities of the coatings were estimated by means of image analysis. The phase compositions of the feedstock powder and coatings were determined by X-ray diffraction (XRD, D/mas-2400, Rigaku, Japan). The oxygen contents in the powder and coatings were measured using ONH-2000 (HRT, Labortechnik, Germany). Each sample measure was repeated 5 times. The microhardness of the coatings was tested under a load of 300 g with a load time of 15 s.

**3. Results and discussions**

**3.1. Coating microstructures**

Coating observations show that the particles can not be deposited to form coatings under the main gas temperature of 300$^o$C. Figure 3 presents the deposition efficiencies of the particles under different main gas temperatures. It is shown that the deposition efficiency of the particles increased from 1.6% to 19.6% with the increase of the main gas temperature from 350$^o$C to 630$^o$C. The deposition efficiency of the particles under the main gas temperature of 630$^o$C is about 13 times that under 350$^o$C. The increase of the deposition efficiency of the particles results from the more intensive deformation due to the higher particle velocity and temperature under the higher main gas temperature, which will be discussed in the next section. Generally speaking, the deposition efficiency of the particles by cold spraying is near to 100% for some non-ferrous metallic materials like copper, aluminum, but not for magnesium. The possible explanation for this relative low deposition efficiency of the particles is the poor deformation ability of Mg particle.

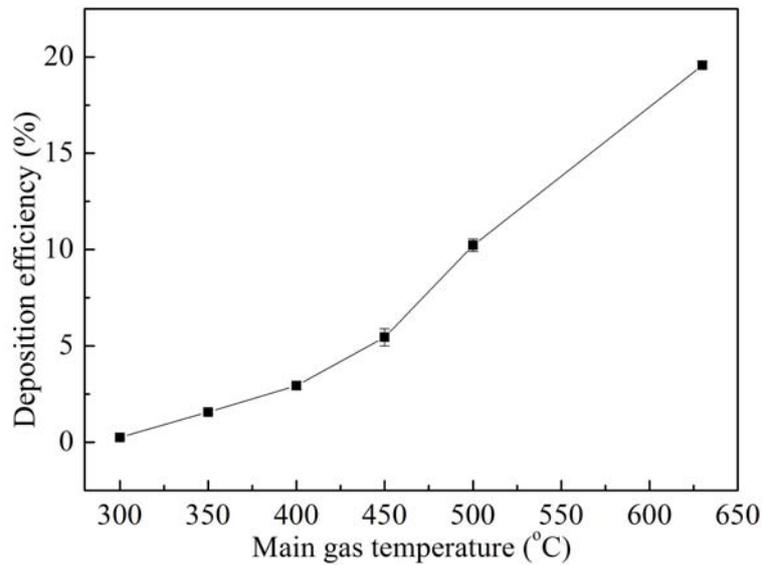

Fig. 3. Effect of main gas temperature on deposition efficiency of particles.

Figure 4 shows the cross-sectional microstructures of the coatings sprayed under the main gas temperature from 350$^{o}$C to 630$^{o}$C. Black zones indicated by arrows in Fig. 4(a) correspond to pores. From Fig. 4(a)-(e), it can be noted that the coating porosity decreased with the increase of the main gas temperature. In order to observe more clearly the role of the main gas temperature, the porosities of the coatings under different main gas temperatures were estimated, and the result is shown in Fig. 5. It can be observed that the porosities of the coatings decreased from 10 vol.% to 1.4 vol.% with the increase of the main gas temperature from 350$^{o}$C to 630$^{o}$C. The sprayed particle can achieve the higher velocity under the higher main gas temperature, thus the deformation extent of the particle increased.

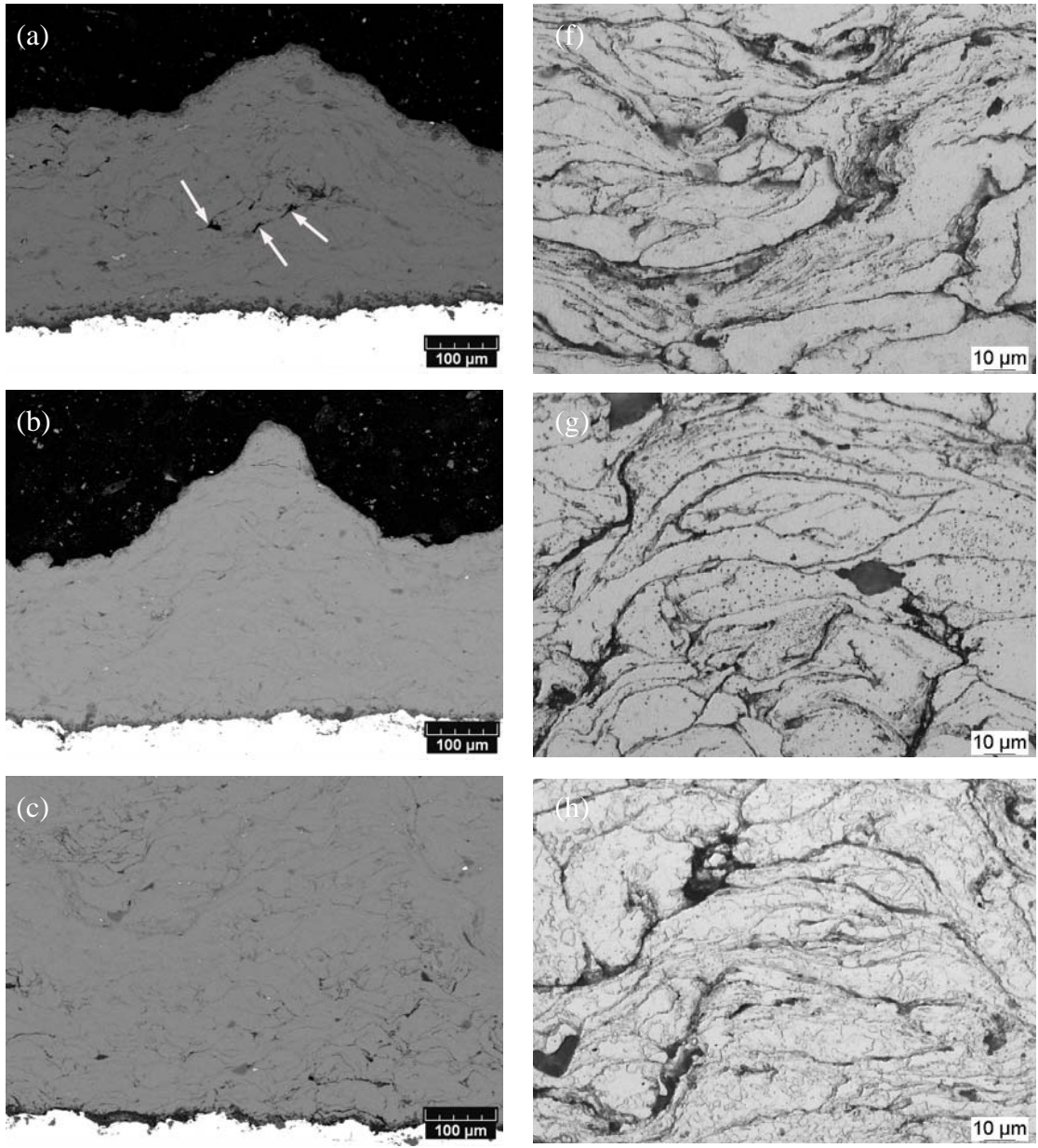

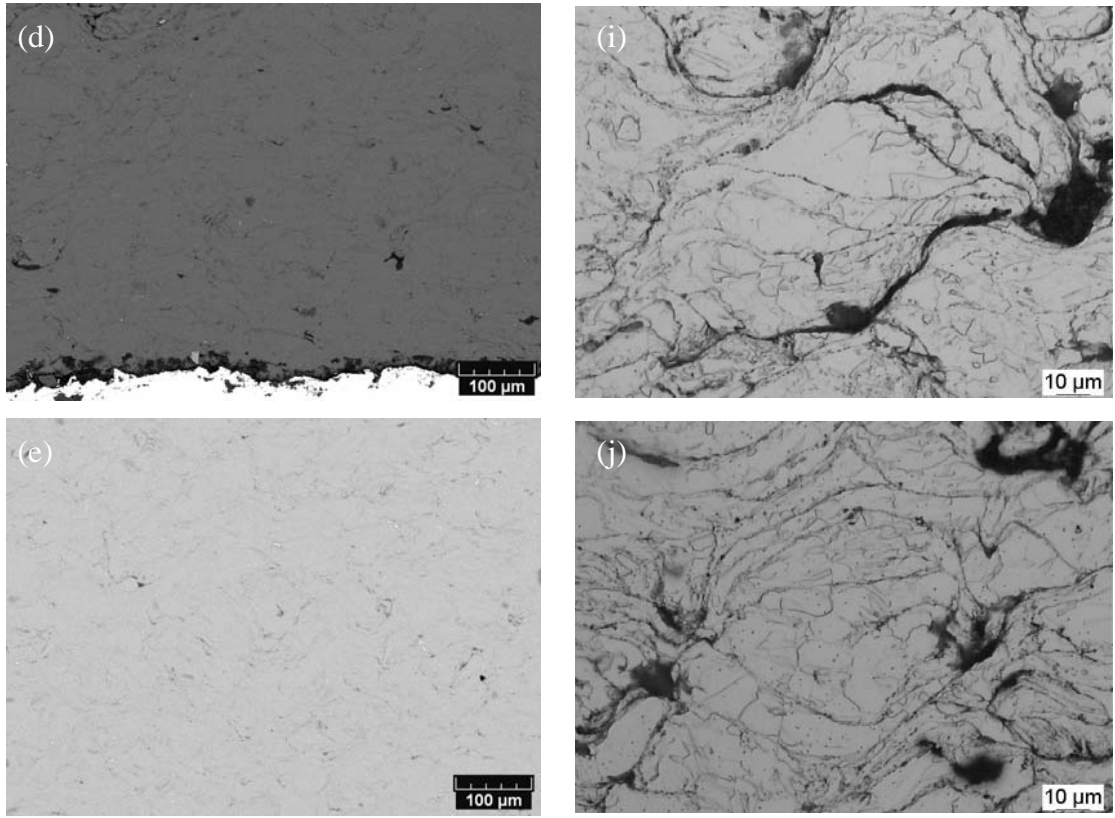

Fig. 4. Cross-sectional microstructure of the coating sprayed under the main gas temperature of (a) 350°C, (f) 350°C, etched, (b) 400°C, (g) 400°C, etched, (c) 450°C, (h) 450°C, etched, (d) 500°C, (i) 500°C, etched, (e) 630°C and (j) 630°C, etched.

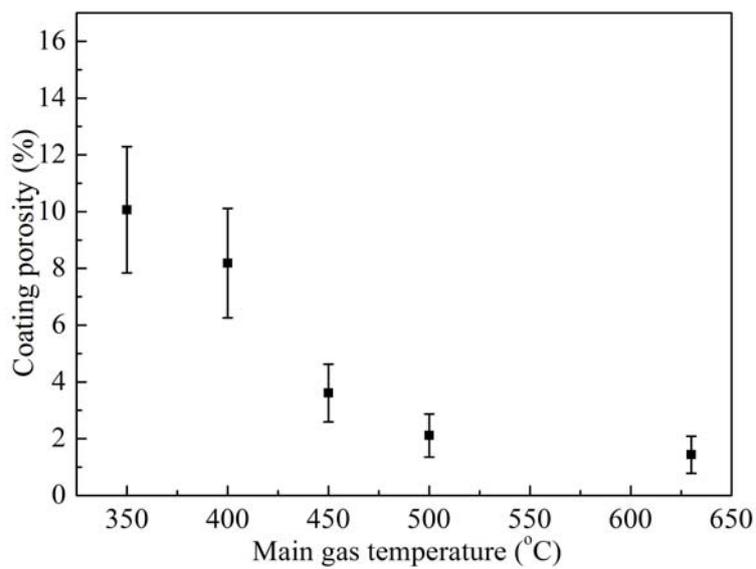

Fig. 5. Effect of main gas temperature on coating porosity.

Figure 4(f)-(j) shows the etched cross-sectional microstructures of coatings sprayed under different main gas temperatures. The lamellar structures can be observed and can be related to the particle deformation during cold spray process. When comparing Fig. 4(h)-(j) with Fig. 1(b), it is found that the grains of the feedstock particles deformed due to the impact process, which were initially equiaxed crystal as shown in Fig. 1(b). However some elongated grains can be found in the coatings as shown in Fig. 4(h)-(j). The elongated grains of the particles in the coatings verify the plastic deformation of the particles during cold spraying.

Figure 6 shows XRD patterns of the feedstock powder and coatings, respectively. It can be observed that only Mg phase is presented in both powder and coatings. It proves that there is no obvious oxidation or deterioration of the particles during cold spraying. By comparing XRD patterns of the powder and coatings, an interesting phenomenon is found. The relative intensity of (0002) crystal face in XRD patterns of the coatings is obviously enhanced compared to that of the powder. This may be due to the texture caused by the plastic deformation of the particles during impact (Chang et al., 2009; Gehrmann et al., 2005; Yu et al., 2009). During impact, the grain would turn to some direction, and thus (0002) crystal face would also turn to this direction. The occurrence of the texture could confirm the plastic deformation of the particles.

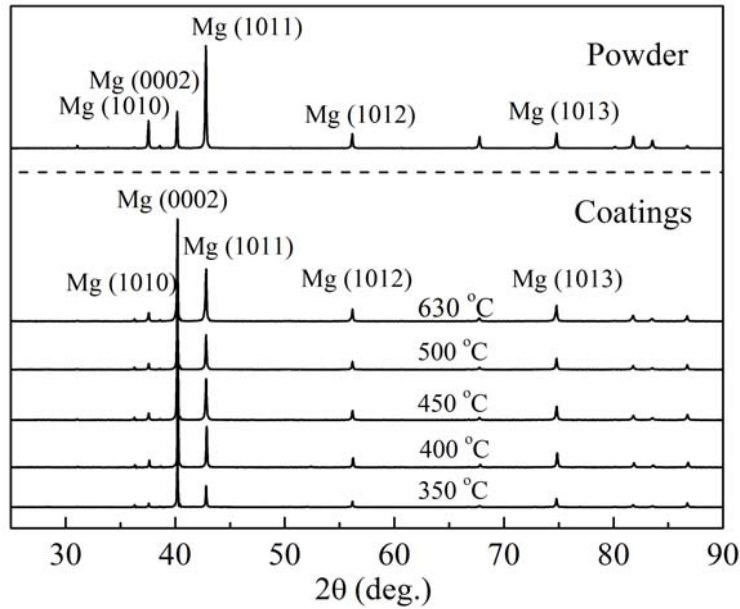

Fig. 6. XRD patterns of the powder and coatings.

Oxygen content test was employed in order to examine the particle oxidation during cold spraying, and the result is shown in Table 1. The feedstock powder was conserved in vacuum, thus the oxygen content keeps a low level which is about 866 ppm. The oxygen content of the as-sprayed coating under 350°C is about 908 ppm, which suggests that the particles did not be oxidized due to the low-temperature characteristic of cold spraying. The main gas temperature utilized in this study is from 350°C to 630°C, and the heating time of the sprayed particles resting in the high temperature flow is very short, about 0.3 ms, thus the particles can effectively avoid oxidization. It is worth noticing that the oxygen content of the coating decreases to 229 ppm as the main gas temperature increases to 630°C, which may benefit from the fragmentation of the oxide film of the particles. During cold spraying, the oxide film of the particles can be broken up, and metallurgical bonding could be achieved. Therefore, the higher particle velocity induced by higher main gas temperature could result in more fragmentation of the oxide film.

Table 1. Oxygen content of the starting powder and coatings under different main gas temperatures.

|  | Oxygen content (ppm) |
|---|---|
| Starting powder | 866±152 |
| Coating sprayed at 350°C | 908±143 |
| Coating sprayedat 630°C | 229±124 |

**3.2. Discussion of the critical velocity of the particles**

The velocity of a sprayed particle prior to impact determines whether the deposition of the sprayed particle or erosion of the substrate occurs during the impact process of sprayed particle. Generally speaking, there exists a critical velocity of the particle for a given material. Only the sprayed particle having a velocity higher than the critical value can be deposited to form a coating. The sprayed particle with a velocity lower than this critical value will result in the shot peening or erosion of the substrate. Of course, if the velocity is too high, the erosion will occur also. The critical velocity of Mg particle was reported as about 800 m/s, which is calculated by a semi-empirical formula (Schmidt et al., 2006). The present experiments show that the coating can not be formed under the main gas temperature of 300°C, therefore it rationally presume that the velocities of all particles under 300°C are lower than the critical value. Because of the direct dependence of the main gas temperature and the particle velocity, more and more particle can achieve a velocity higher than the critical value if the main gas temperature increases, thus the deposition efficiencies of the particles increased. Figure 7 shows the modeled evolution of in-flight particle velocity as a function of the particle size under different main gas temperatures calculated by FLUENT. It is shown that a smaller particle can achieve a higher particle velocity under the same main gas temperature, and the particle velocity increases with the increase of the main gas temperature. No coating can be deposited at 300°C with the highest particle velocity of 653 m/s. Meanwhile the coating can be deposited at 350°C with the highest particle velocity of 677 m/s. Therefore, the critical velocity of Mg particle is estimated in the range of 653 m/s to 677 m/s.

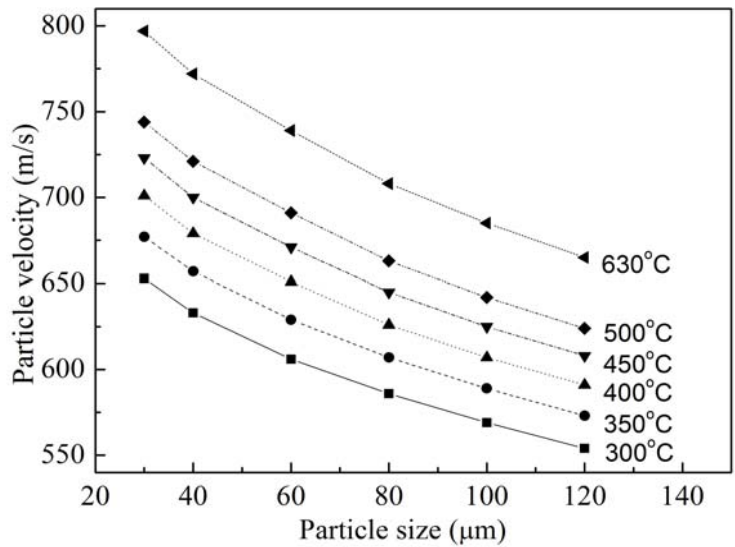

Fig. 7. Effect of particle size on particle velocity under different main gas temperatures.

3.3. Bonding mechanism of the particles

Figure 8 displays the fracture morphologies of the coatings under different main gas temperatures. The fracture morphology of the coating under the main gas temperatures of 350°C given in Fig. 8(a) is clearly different from others. There is no obvious phenomenon of squeezed particles. The black arrows in Fig. 8(b) correspond to the impact direction of two particles. The subsequent particle impacted onto the deposited particle and the interfaces of these two particles could be found clearly to contact closely with each other. At 500°C and 630°C, an obvious laminar manner could be observed in Fig. 8(d) and (e). These apparent evidences of the plastic deformation not only result from the high particle velocity, but also from the higher particle temperature. The effect of the particle velocity on the plastic deformation of particles can be observed in many articles. For the materials with a number of slip systems is relevant to the temperature, the effect of the particle temperature on the plastic deformation of the particles should not be ignored. When the temperature of Mg particle is below to 225$^{o}$C, it has only one glide $\{0001\}$ $<11\bar{2}0>$ in basal plane and one twinning system $\{10\bar{1}2\}$ $<10\bar{1}1>$ in pyramidal plane. While the particle temperature exceeds 225$^{o}$C, another new glide system $\{10\bar{1}0\}$ $<11\bar{2}0>$ in prismatic plane could slip (Couret and Caillard, 1985). The particle temperature calculated by Fluent with the average diameter of 63.2 µm under the main gas temperatures of 500$^{o}$C and 630$^{o}$C is 260$^{o}$C and 330$^{o}$C, respectively, which exceeds 225$^{o}$C. Consequently, the plastic deformability of Mg particle enhances. Moreover, no evidence of melting and metallurgical bonding could be observed. Thus the bonding mechanism of the coating is mainly the mechanical interlocking. It can be estimated that the bond of the coating and the substrate or between the particles is weak.

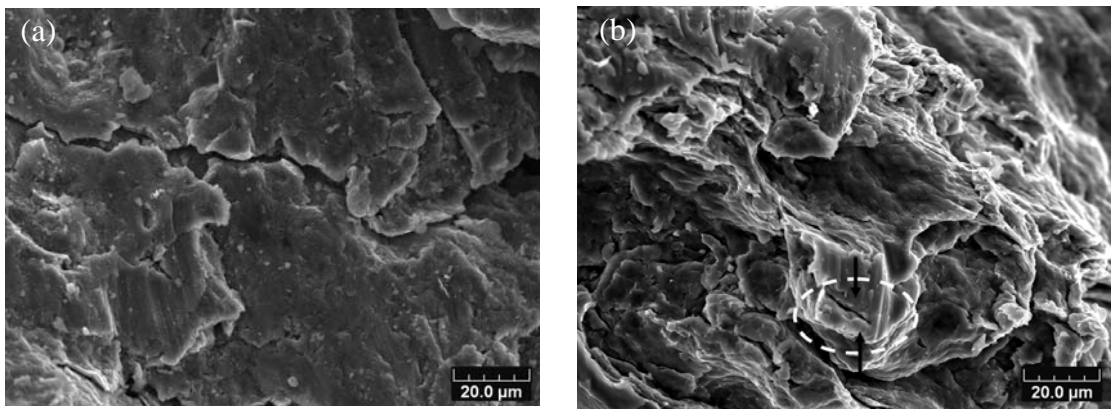

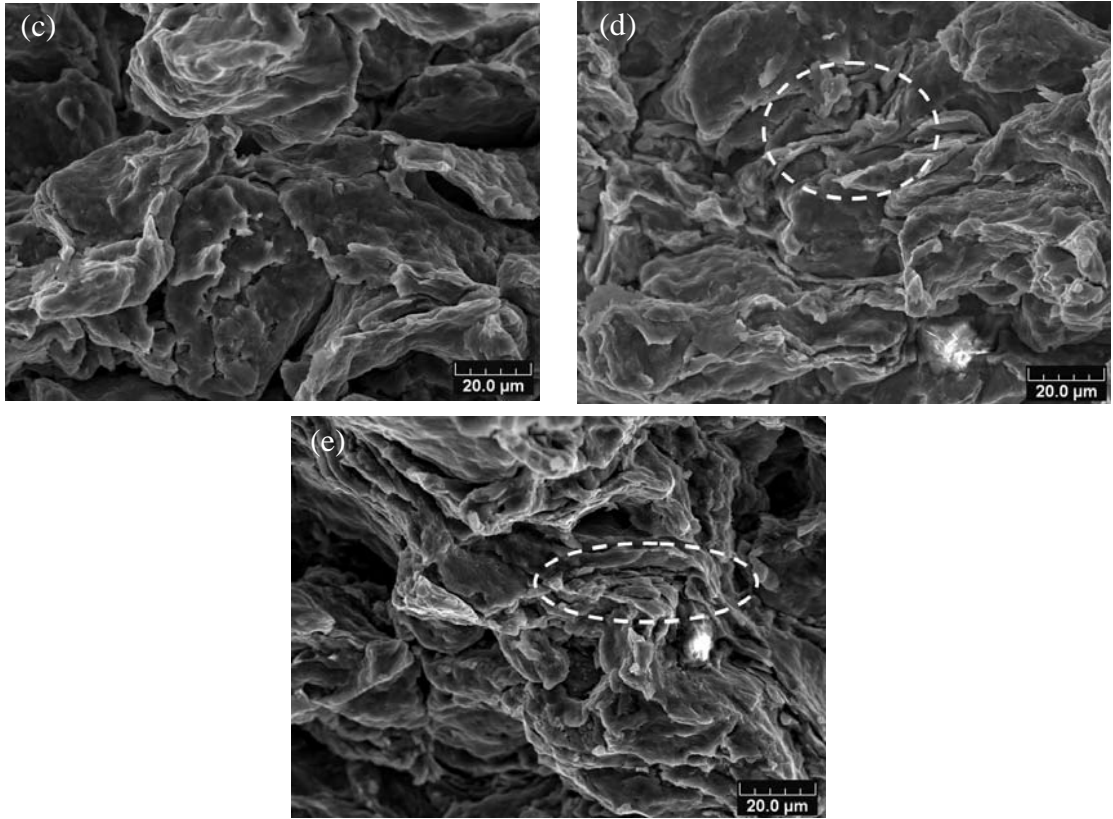

Fig. 8. Fracture morphology of the coating sprayed under the main gas temperature of

(a) 350°C, (b) 400°C, (c) 450°C, (d) 500°C and (e) 630°C.

3.3. Vickers microhardness of coatings

Figure 9 presents the microhardness of the coatings prepared under different main gas temperatures. The microhardness of the coatings increased as the main gas temperature increased from 350°C to 450°C, because the coatings became denser at higher main gas temperature. This result is in agreement with the porosities of the coatings as shown in Fig. 5. The porosities of the coatings became less with the increase of the main gas temperature ranging from 350°C to 450°C. When the main gas temperature was higher than 450°C, the microhardness of the coatings varied slightly. The microhardness of the coatings under the main gas temperature from 450° to 630°C was about 38 $HV_{0.3}$. This is close to the microhardness of bulk Mg prepared by casting which was about 43 HV (Hassan and Gupta, 2002; Saravanan and Surappa, 2000).

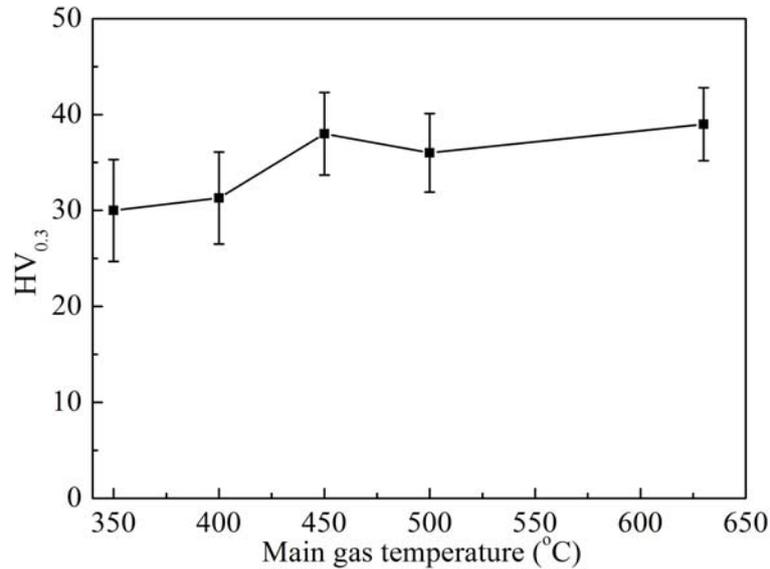

Fig. 9. Effect of main gas temperature on microhardness of coatings sprayed under different main gas temperatures.

## 4. Conclusions

Based on the results obtained in this study, the following conclusions can be drawn.

1. Dense Mg coatings were successfully deposited by cold spraying.

2. No coating can be formed with the lowest main gas temperature of 300°C. Coatings could be built for the main gas temperature higher than 350°C. According to the calculation, the critical velocity of Mg particles was in the range of 653 m/s to 677 m/s.

3. The coating porosity decreased with the increase of main gas temperatures, and the oxygen content of the coatings didn't increase compared with that of the feedstock powder. The coating microhardness increased with the increase of main gas temperatures.

4. The bonding mechanism of Mg particles was mainly the intensive plastic deformation and mechanical interlocking.